\documentclass[manuscript]{aastex}

\newcommand{\Lya}{\mbox{Ly$\alpha$ }}

\shorttitle{Constraining $H_0$ from \Lya and BAO}
\shortauthors{Busti et al.}

\begin{document}

\title{Constraining $H_0$ from Lyman-$\alpha$ Forest and Baryon Acoustic Oscillations}

\author{V. C. Busti}
\affil{Departamento de Astronomia, Universidade de S\~ao Paulo, USP,
    S\~ao Paulo, SP, Brazil 05508-090}
\email{vcbusti@astro.iag.usp.br}

\author{ R. N. Guimar\~ aes}
\affil{Programa de Modelagem Computacional, SENAI - Cimatec, Salvador, BA, Brazil 41650-010}
\email{rguimara@eso.org}

\and

\author{J. A. S. Lima}
\affil{Departamento de Astronomia, Universidade de S\~ao Paulo, USP,
    S\~ao Paulo, SP, Brazil 05508-090}
\email{limajas@astro.iag.usp.br}

\begin{abstract}
A new method is proposed to measure the Hubble constant $H_0$ through the mean transmitted flux observed from high redshift quasars. A 
semi-analytical model for the cosmological-independent volume density distribution function is adopted which allows one to obtain constraints 
over the cosmological parameters once a moderate knowlegde of the InterGalactic Medium (IGM) parameters is assumed. By assuming  a flat $\Lambda$CDM cosmology,  
we show that such method alone cannot provide good constraints on the pair of free parameters ($h, \Omega_m$). However, it is possible  possible to break the degeneracy 
on the mass density parameter by applying a joint analysis involving the baryon acoustic oscillations (BAOs). Our analysis based on two different samples of Lyman-$\alpha$ forest resctricts ($h, \Omega_m$) on the intervals $0.58 \leq h \leq 0.91$ and $0.215 \leq \Omega_m \leq 0.245$ ($1\sigma$). Although the constraints are weaker comparatively to other estimates,  we point out that with a bigger sample and a better knowledge 
of the IGM this method may present competitive results to measure the Hubble constant 
independently of the cosmic distance ladder.
\end{abstract}



\section{Introduction}

Measurements of the Hubble constant $H_0$ are vital to the establishment of a cosmic concordance model in cosmology. 
Such constant plays a role in most of  cosmic calculations, as the age of the Universe, its size and energy density, primordial nucleosynthesis, physical 
distances to objects and so on. Its importance must even increase in the next decade because new missions
and observational projects are being designed to provide accurate measurements of the Hubble constant thereby also increasing the precision in the other cosmological parameters. 

There are several ways to determine $H_0$ by using low and high redshift sources. The most commom methods are based on the period-luminosity relation seen in Cepheids, tip of the red giant branch and Supernovae observed in the local Universe (Freedman \& Madore 2010). However, some alternative procedures like the  Sunyaev--Zel'dovich effect combined with X-ray emission from clusters allow a measurement of $H_0$ from high-redshift objects  have also been discussed (Cunha et al. 2007, Holanda et al. 2010).
Although considering that some errors recently reported in the literature are small (e.g. Riess et al. 2009), discrepancies among different measurements as the value given by \cite{Sandage06} show that independent estimates are needed in order to achieve a reliable value for $H_0$ not plagued by systematic errors from astrophysical environments, as well as  from calibrations associated to the cosmic distance ladder. In this concern, the importance and cosmological interest on different estimates of $H_0$ independent of the 
distance ladder has also been discussed by many authors (see Jackson 2007 and Refs. therein for a recent review).

A possible alternative procedure to measure $H_0$ is to consider the Lyman-$\alpha$ (\Lya) forest data that probes the low density intergalatic medium (IGM) over a unique range of redshifts and environments  as seen  in the spectra from high redshift quasars. The observed mean transmitted flux depends on the local optical depth, which in turn depends on the expansion rate of the Universe $H(z)$ and 
some properties characterizing the intervening medium along our line of sight to any quasar. In principle, with a moderate knowledge of the IGM properties, as the hydrogen photoionization rate and mean temperature, the values of some cosmological parameters including  the Hubble constant can be constrained. 

In the last decade, several studies based on \Lya forests as cosmological tools have been performed (Macdonald et al. 2000, Weinberg et al. 2003, Wyithe et al. 2008, Viel et al. 2009). Some works use the flux power spectrum of the \Lya forest to infer the cosmological parameters in  two ways: inversion of the flux power spectrum to get the underlying dark matter power spectrum \citep{croft_1998,croft_1999,hui1999,nusser_1999}, or to use the power spectrum  directly \citep{mcdonald_2000,mcdonald_2005,mandelbaum_2003}. 

On the other hand, a simpler approach is to adopt a semi-analytical model to describe the IGM as proposed by \cite{Miralda_2000}, who  derived a fitting formula for the volume density distribution function in agreement with simulations which provided similar results for different numerical methods and different cosmologies \citep{rauch_1997}. Therefore, a theoretical mean transmitted flux is obtained and compared to observational data in order to constrain the cosmological parameters. 

In this letter, by assuming  a flat $\Lambda$CDM cosmology, we apply the  above described procedure for deriving new constraints on
the Hubble constant $H_0$ and the matter density parameter $\Omega_m$. By considering two independent samples of \Lya forest as compiled by Bergeron et al. (2004) and  Guimar\~aes et al. (2007), respectively,   two distinct statistical analyses are performed. Firstly, we consider only the \Lya forest data set whereas  the second approach involves a
joint analysis combining the \Lya forest data with the SDSS measurements of the baryon acoustic peak from  the clustering of luminous red galaxies \citep{Eisenstein05}.  As we shall see, the degeneracy on the free parameters defining the ($h, \Omega_m$) plane is naturally broken when the \Lya data are combined with the standard ruler as given  by the baryon acoustic oscilations thereby providing a new independent method to estimate the Hubble parameter.

\section{The \Lya Forest Samples}

As remarked earlier, in the present analysis we consider two different samples. The first one consists of 18 high-resolution high signal-to-noise ratio (S/N) spectra. The spectra were obtained with the Ultra-Violet and 
Visible Echelle Spectrograph (UVES) mounted on the ESO KUEYEN 8.2 m telescope at the Paranal observatory for the ESO-VLT Large Programme (LP) 'Cosmological evolution of the Inter Galactic Medium' (PI Jacqueline Bergeron; Bergeron et al. 2004).  The spectra were taken from the European Southern Observatory archive and are publicly available. This  programme has been devised to gather a homogeneous sample of echelle spectra of 18 QSOs, with uniform spectral coverage, resolution and signal-to-noise ratio suitable for studying the IGM. The spectra have a signal-to-noise ratio of 40 to 80 per pixel and a spectral  resolution $\lambda / \Delta\lambda \sim 45 000$ in the \Lya forest region. Details of the procedure used  for data reduction can be found in \cite{chand_2004} and \cite{aracil_2004}. 

The second sample consists of 45 quasars obtained by \cite{rodney}, where the spectra were measured with the ESI \citep{sheinis2002} mounted on the Keck II 10m telescope. The spectral resolution is $R \sim 4300$ with a signal-to-noise ratio usually larger than 15 per 10 km s$^{-1}$ pixel. This sample has also been studied Prochaska and collaborators (2003a,b) in their discussions related to the existence of Damped \Lya systems. In the present analysis we have used the \Lya forest over the rest-wavelenght range 1070-1170~\AA~ to avoid contamination by the proximity effect close to the \Lya  emission line and possible OVI associated absorbers. We also carefully avoided regions flagged because of damped/sub-damped Ly$\alpha$ absorption lines. We divided each spectrum in bins of length $100$~\AA~ in the observed frame, corresponding to about $\triangle z = 0.08$, for the last bin we have used a 
lenght of $500$~\AA~ to keep the statistical homogeneity. The mean transmitted flux was calculated as the mean flux over all pixels in a bin. At each redshift, we then averaged 
the transmitted fluxes over all spectra covering this redshift. 

\begin{figure}
\plotone{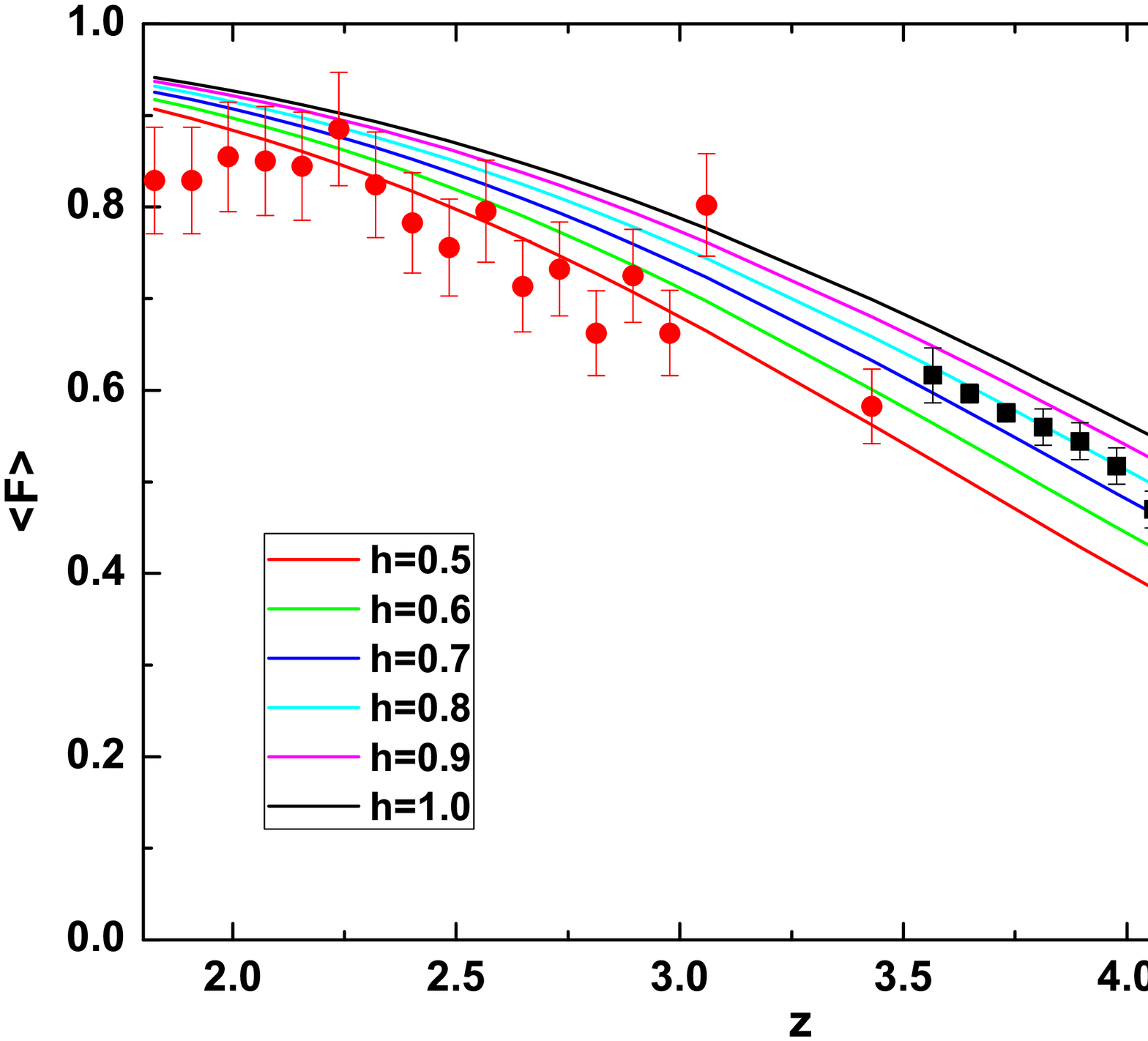}
\caption{Mean transmitted flux as a function of redshift for $\Omega_{m}=0.23$, $\Gamma = 0.8\times 10^{-12}$ s$^{-1}$, $T_0=2.5 \times 10^4$ K and some selected values of 
the $h$ parameter. The data points correspond to the \Lya forest measurements obtained by this work (red circles) and  \citet{rodney} (black squares).}\label{Fig1}
\end{figure}

In  Fig~\ref{Fig1}, we display the  mean transmitted flux for the two samples.  Notice that all errors were estimated as the standard deviation of the mean values divided by the square root of the number of spectra. The corresponding scatter in the transmitted 
flux cannot be explained by photon noise or by uncertainties in the continuum thereby suggesting that errors are dominated by cosmic variance.

\section{Basic Equations}

The basic measurement obtained from the \Lya forest is the mean transmitted flux at redshift $z$ defined as

\begin{equation}
\langle F \rangle (z) \equiv \langle \exp{(-\tau(z))} \rangle,
\end{equation}
where  $\tau(z)$ is the local optical depth and angle brackets denote an average over the line of sight. Following standard lines  \citep{huignedin1997}, throughout this paper photoionization equilibrium  and a power-law temperature-density relation, $T=T_{0}(1+\delta)^{\beta}$, for the low-density IGM will be assumed, where $\delta$ stands for the 
local overdensity and $T_{0}$ for the IGM temperature at mean density ($\delta=0$). The local optical depth is given by \citep{peebles1993,padmanabhan2002,faucher_apjl_2008}

\begin{equation}
\tau(z) = A(z) (1+\delta)^{2-0.75\beta},
\end{equation}
with

\begin{equation}
A(z)  \equiv
\frac{\pi e^{2} f_{Ly\alpha}}{m_{e} \nu_{Ly\alpha}}
\left( \frac{\rho_{\rm crit} \Omega_{b}}{m_{p}} \right)^{2}
\frac{X(X+0.5Y)}{H(z)} 
\frac{R_{0} T_{0}^{-0.75}}{\Gamma}
(1 + z)^{6}.
%\end{array}
\end{equation}

In the expression above, $f_{Ly\alpha}$ is the oscillator strength of the \Lya~transition, $\nu_{Ly\alpha}$ is its frequency, $m_e$ and $m_p$ are the electron and proton 
masses respectively, $\Omega_b$ is the baryon density parameter, $H(z)$ is the Hubble parameter, $\rho_{\rm crit}$ is the critical density, $X$ and $Y$ are the mass fractions 
of hydrogen and helium respectively taken to be 0.75 and 0.25 \citep{burles_apj_2001}, $R_{0}=4.2\times10^{-13}~{\rm cm}^{3}~{\rm s}^{-1}/(10^{4}~{\rm K})^{-0.75}$ and $\Gamma$ is 
the hydrogen photoionization rate. 

The mean transmitted flux is obtained by integrating the local optical depth through a volume density distribution function for the gas $\Delta \equiv 1 + \delta$,
\begin{equation}
\langle F \rangle(z) =
\int_{0}^{\infty}
d\Delta
P(\Delta; z)
\exp{[-\tau(z)]}.
\end{equation}

As is widely known, \cite{Miralda_2000} derived an approximate analytical functional form for the distribution function given by
\begin{equation}
P(\Delta;z) =
A \exp{\left[
-\frac{(\Delta^{-2/3} - C_{0})^{2}}{2(2\delta_{0}/3)^{2}}
\right]}\Delta^{-b},
\end{equation}
where the parameters $A$ and $C_{0}$ are derived by requiring the total volume and mass to be normalized to unity. \cite{Miralda_2000} extrapolated the distribution 
function and obtained $\delta_{0}=7.61/(1+z)$ with an accuracy better than 1\% from fits to a numerical simulation at $z=$2, 3, and 4 from \cite{miralda1996}. In order to apply this 
formalism to our data, the values of $b$ were derived from a cubic interpolation from the values in the simulation. Applications of this distribution function to constrain cosmological 
parameters is well justified since simulations with different numerical methods and different cosmologies provided very similar results \citep{rauch_1997}.

From now on a flat $\Lambda$CDM cosmology will be assumed. In this case, the Hubble parameter is given by

\begin{equation}
H(z) = H_0\left[\Omega_m(1 + z)^{3} + (1 -\Omega_m)\right]^{1/2},
\label{eq2}
\end{equation}
where we will adopt the convention $h=H_0/100$ km s$^{-1}$ Mpc$^{-1}$, which is the Hubble constant normalized in units of $100$ km s$^{-1}$ Mpc$^{-1}$.

The effects of $H_0$ in the mean transmitted flux are shown in Fig. \ref{Fig1} along with our observational data (red circles) and \cite{rodney} (black squares). We see that with 
our sample we obtain a scattered data set comparatively to the sample used by \cite{rodney}, which is related to the sample size. It is also evident that our data cover most of 
values for the $h$ parameter, so we do not expect tight constraints from the \Lya data set alone.

\section{Analyses and Results}

Let us now perform a statistical analysis to find the constraints on the cosmological parameters. The full set of parameters are represented by $\mathbf{p} \equiv (h,\Omega_{m},\Gamma,T_0,\Omega_{b}h^2,\beta)$. We fix the value of $\Omega_{b}h^2=0.0218$ using the latest observations of deuterium \citep{deuterium} from the Big Bang Nucleosynthesis \citep{bbn1}. An early reionization model is also considered which gives $\beta=0.62$ \citep{huignedin1997}. The posterior probability of the parameters $P(\mathbf{p}|d)$ given the data $d$ is

\begin{equation}
P(\mathbf{p}|d)=\frac{P(d|\mathbf{p})P(\mathbf{p})}{P(d)},
\end{equation}
where $P(d)$ is a normalization constant, $P(\mathbf{p})$ is the prior over the parameters and $P(d|\mathbf{p}) \propto e^{-\chi^{2}/2}$ is the likelihood with the usual definition, for the \Lya forest data, of

\begin{equation}
\chi^{2}_{Ly\alpha} = \sum_i {( \langle F_{th} \rangle(z_i;\mathbf{p})-
\langle F \rangle(z_i))^2 \over \sigma_{\langle F \rangle(z_i)}^2},
\end{equation}
where $\langle F_{th} \rangle(z_i;\mathbf{p})$ is the theoretical mean transmitted flux, $\langle F \rangle(z_i)$ is the observational mean transmitted flux and $\sigma_{\langle F \rangle(z_i)}$ is its respective uncertainty. We treat the combination $\Gamma T_{0}^{0.75}$ as a nuisance parameter with a flat prior. The ranges chosen are $T_0=[2,2.5] \times 10^{4}$ K from estimates of \cite{zaldarriaga2001} and $\Gamma=[0.8,1] \times 10^{-12}$ s$^{-1}$ which cover some measurements reported in the literature \citep{rauch_1997,mcdonald2001,meiksin2004,tytler2004,bolton_2005,kirkman2005}, although in disagreement with the values obtained by \cite{faucher_apjl_2008,faucher_apj_2008}, who used a redshift-dependent relation for $T_0$ not favoured by our data.

In what follows, we first  consider the \Lya forest data separately, and, further, we present a joint analysis including the
BAO signature extracted from the SDSS catalog (Eisenstein et al. 2005).

\subsection{Limits from \Lya Forest data set}

\begin{figure}
\plotone{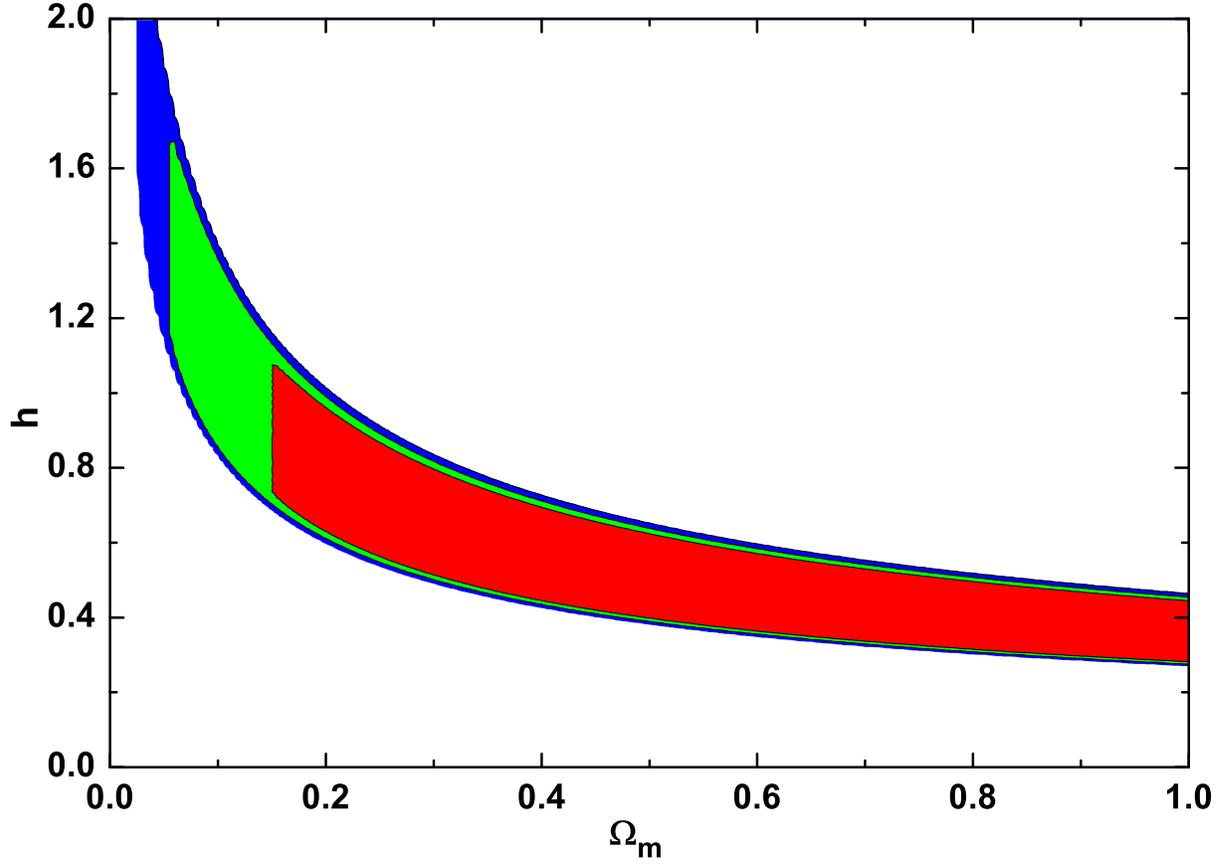}
\caption{Confidence regions ($68.3$\%, $95.4$\% and
$99.7$\%) in the $(\Omega_{m}, h)$ plane provided by \Lya forest data. The basic conclusion is that \Lya forest alone cannot constrain the pair of parameters.}\label{Fig2}
\end{figure}

In Fig. \ref{Fig2} we display the results of the statistical analysis performed with the \Lya forest data. We see that the data do not provide good constraints to both parameters. Several sanity checks were performed in order to determine the values fixed in the statistical analysis. In particular, we have seen that different values for $\beta$, different intervals for $\Gamma T_{0}^{-0.75}$ in the marginalization, or even a redshift-dependence for $T_0$, have resulted in poorer fits compared to what is shown in figure \ref{Fig2}.  In addition, even considering that  cosmological constraints are affected by the IGM physical parameters the opposite vision is also true, that is,  different cosmologies may also provide results not compatible for the IGM parameters. In this context, it is recommended to fit cosmological and IGM parameters together in order to obtain more trustful results.

Despite the considerations discussed above, we see from Fig. \ref{Fig2} that the slope in the $(h,\Omega_m)$ plane suggests that a joint analysis with an independent  test constraining  only $\Omega_m$ could provide interesting limits to the Hubble parameter. Thus, a joint analysis with the BAO data is presented in the next subsection.

\subsection{\Lya Forest and BAO: A  Joint Analysis }

In order to achieve better constraints on the cosmic parameters  we apply a joint analysis involving \Lya forest data and baryon acoustic oscillations data obtained from 46748 luminous red galaxies selected from the SDSS Main Sample. The BAO scale can be represented by the parameter \citep{Eisenstein05}

\begin{eqnarray}
 {\cal{A}} \equiv {\Omega_{\rm{m}}^{1/2} \over
 {{\cal{H}}(z_{\rm{*}})}^{1/3}}\left[\frac{1}{z_{\rm{*}}}
 \Gamma(z_*)\right]^{2/3}  = 0.469 \pm 0.017,
\label{eqA}
\end{eqnarray}
where $z_{\rm{*}} = 0.35$ is the redshift at which the acoustic
scale has been measured, ${\cal{H}}(z)$ is $H(z)/H_0$ and $\Gamma(z_*)$ is the dimensionless
comoving distance to $z_*$.

From equation (\ref{eqA}) it is seen that the BAO scale is independent of $h$ thereby yielding constraints only for the matter density parameter. The statistical analysis is performed with $\chi^2 = \chi^{2}_{Ly\alpha}+\chi^{2}_{BAO}$, where

\begin{equation}
\chi^{2}_{BAO}=\left(\frac{{\cal{A}}-0.469}{0.017}\right)^2.
\end{equation}

\begin{figure}
\plotone{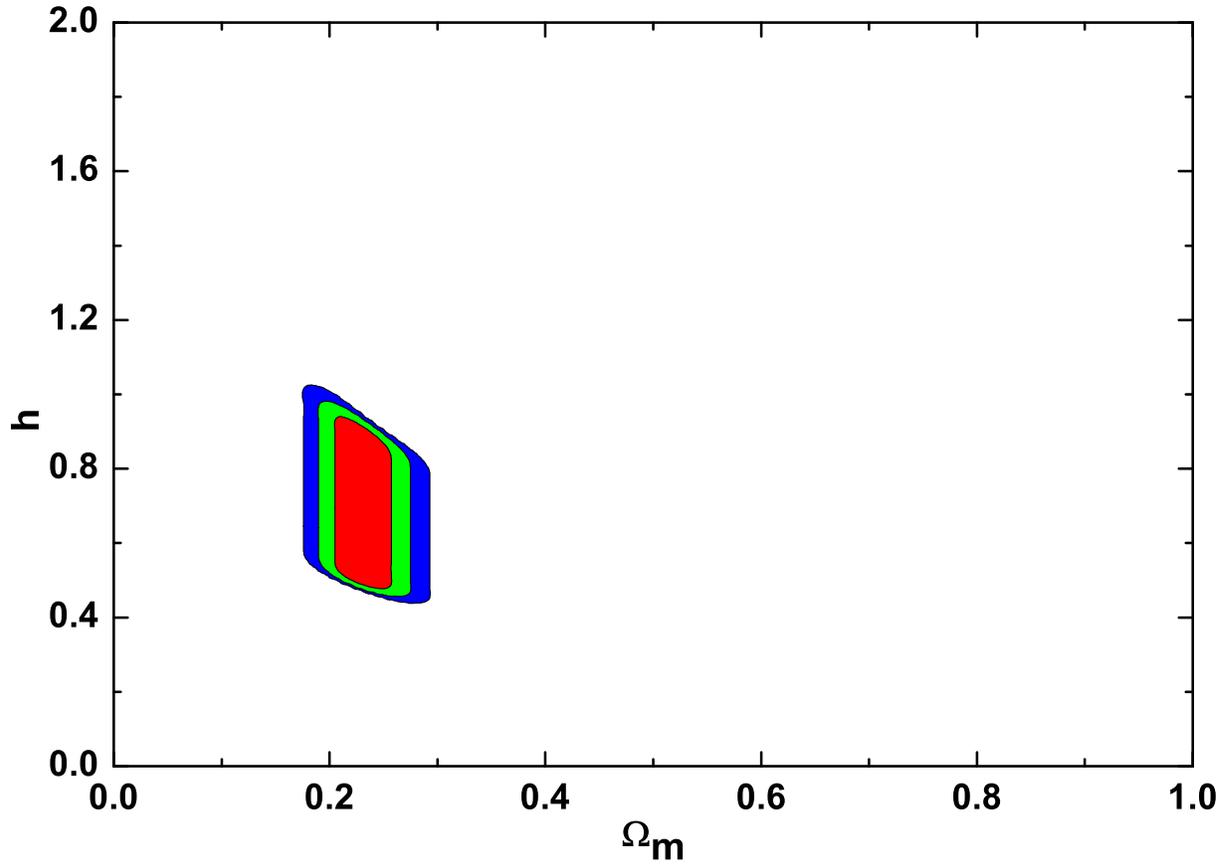} 
\caption{Contours in the $\Omega_m - h$ plane provided by 
the joint analysis combinig \Lya Forest and BAO. As before, the contours correspond to
$68.3$\%, $95.4$\% and $99.7$\% confidence levels. Note that the best-fit
model converges to $h = 0.67$ and $\Omega_m=0.23$. These results should be compared with the ones presented in Figure 2.}\label{Fig3}
\end{figure}

In Fig. \ref{Fig3}, we show the limits on the pair of parameters $(h,\Omega_m)$ obtained from our joint analysis involving \Lya forest data and BAO. Within 68.3\% confidence level, we have obtained $0.215 \leq \Omega_m \leq 0.245$ and $0.58 \leq h \leq 0.91$. The value for $\Omega_m$ is lower than what was found by other observations \citep[e.g.][]{amanullah_2010}, but it is likely that the difference is due to systematic errors not taken into account in our analysis. 

In Tab. \ref{Tab1}, we show some recent measurements of $H_0$ using different techniques and data. Although existing a tension among different estimates \citep[e.g.][]{Sandage06,dark}, our approach do not provide stringent limits which would allow a decision about the correct  value of $H_0$. The interesting aspect is that the method discussed here provides an independent estimate which is in agreement with the values obtained adopting different approaches (see Table \ref{Tab1}).

\section{Conclusions}

In this work we have used a cosmological-independent semi-analytical model to describe the IGM and data from \Lya forest and baryon acoustic oscillations to constrain cosmological parameters. Limits on the Hubble constant $H_0$ and the matter density parameter $\Omega_m$ were derived by assuming a flat $\Lambda$CDM cosmology.  

By applying a statistical analysis using only the \Lya forest data we did not obtain good constraints to both parameters. However, when we applied a joint analysis involving the \Lya forest data and BAO interesting limits were found, being the parameters restricted to the intervals $0.215 \leq \Omega_m \leq 0.245$ and $0.58 \leq h \leq 0.91$ within 68.3\% confidence level. These results are in agreement with recent measurements reported in the literature (see Tab. \ref{Tab1}), but are weaker due to the limited sample and our poor knowledge of the IGM. We expect better constraints to the Hubble constant in the near future by using a bigger sample and a deeper understanding of the IGM, which will turn this technique competitive with other estimates with the advantage of being independent of a cosmic distance ladder.

\acknowledgments

The authors are grateful to J. V. Cunha and J. F. Jesus  for
helpful discussions. VCB is supported by CNPq and RNG is partially supported FAPESB. 
JASL is partially supported by CNPq and FAPESP (No. 04/13668-0).

\clearpage

%% the file, uncomment the \plotone, \plottwo, and
%% If you need a layout that cannot be achieved with \plotone or
%% \plottwo, you can invoke the graphicx package directly with the
%% \includegraphics command or use \plotfiddle. For more information,
%% please see the tutorial on "Using Electronic Art with AASTeX" in the
%% documentation section at the AASTeX Web site,
%% http://www.journals.uchicago.edu/AAS/AASTeX.
%%
%% The examples below also include sample markup for submission of
%% supplemental electronic materials. As always, be sure to check
%% the instructions to authors for the journal you are submitting to
%% for specific submissions guidelines as they vary from

%% This example uses \plotone to include an EPS file scaled to
%% 80% of its natural size with \epsscale. Its caption
%% has been written to indicate that additional figure parts will be
%% available in the electronic journal.

\clearpage

\begin{table*}
\begin{center}
\footnotesize \small
\caption{Limits to $H_0$}
% ***********************************************
\label{Tab1} \vspace{1pc}
\begin{tabular}{l c c }
\hline
Method & Reference & h\\
\hline
Cepheid Variables & \cite{WFred01} (HST Project) & $0.72 \pm 0.08$  \\
Age Redshift & \cite{jimenez2003} (SDSS) & $0.69\pm 0.12$  \\
SNe Ia/Cepheid & \cite{Sandage06} & $0.62 \pm 0.013$(rand.)$\pm 0.05$(syst.)  \\
SZE+BAO & \cite{cunha_2007} & $0.74^{+0.04}_{-0.03}$  \\
Old Galaxies + BAO & \cite{lima2009} & $0.71 \pm 0.04$  \\
SNe Ia/Cepheid & \cite{riess2009} & $0.742 \pm 0.036$  \\
CMB & \cite{komatsu2010} (WMAP) & $0.710 \pm 0.025$ \\
SZE+BAO & \cite{holanda_2010} & $0.732^{+0.043}_{-0.037}$  \\
Time-delay lenses & \cite{dark} & $0.76 \pm 0.03 $  \\
{\bf \Lya + BAO} & {\bf This work} &  $0.67^{+0.24}_{-0.09}$ \\
\hline\\
\end{tabular}
\end{center}
\end{table*}

%% Tables may also be prepared as separate files. See the accompanying
%% sample file table.tex for an example of an external table file.
%% To include an external file in your main document, use the \input
%% command. Uncomment the line below to include table.tex in this
%% sample file. (Note that you will need to comment out the \documentclass,
%% \begin{document}, and \end{document} commands from table.tex if you want
%% to include it in this document.)


%% The following command ends your manuscript. LaTeX will ignore any text

\end{document}